\documentstyle[epsfig,longtable]{aipproc}

\newbox\grsign \setbox\grsign=\hbox{$>$}
\newdimen\grdimen \grdimen=\ht\grsign
\newbox\laxbox \newbox\gaxbox
\setbox\gaxbox=\hbox{\raise.5ex\hbox{$>$}\llap
     {\lower.5ex\hbox{$\sim$}}}\ht1=\grdimen\dp1=0pt
\setbox\laxbox=\hbox{\raise.5ex\hbox{$<$}\llap
     {\lower.5ex\hbox{$\sim$}}}\ht2=\grdimen\dp2=0pt
\def\gax{\mathrel{\copy\gaxbox}}
\def\lax{\mathrel{\copy\laxbox}}
\def\lta{\lax}
\def\gta{\gax}

\begin{document}

\title{DISK INSTABILITY MODELS}
 
\author{Jean--Pierre Lasota$^*$ and Jean--Marie Hameury$^{\dagger}$}
\address{$^*$UPR 176 du CNRS; DARC Observatoire de Paris, Meudon, France\\
$^{\dagger}$UMR 7550 du CNRS; Observatoire de Strasbourg, Strasbourg, France}

\maketitle

\begin{abstract}
We review various aspects of disk instability models. We discuss 
problems and difficulties and present ways that have been suggested to solve
them.
\end{abstract}

\section*{Introduction}

Almost everyone interested in the subject would agree with the
statement that the thermal--viscous disk instability model (DIM)
describes both dwarf nova (DN) outbursts and soft X-ray transients. If
one asked, however, what is meant by the thermal--viscous DIM, the answer
would be far from unanimous; and if one asked for examples of systems
correctly described by the DIM, there would be no answer at all. One
or two examples would we given but a confrontation of the DIM
predictions with observations would show them to be too optimistic.

What we usually mean when we say that the DIM is the model
describing DNs and SXTs is that it is a general idea (we
resisted the temptation to call it a `paradigm') according to which
both DN outbursts and SXTs are triggered by an accretion disk
instability due to an abrupt change in opacities at temperatures at
which hydrogen is partially ionized. All versions of the DIM have this
ingredient. They differ in assumptions about viscosity, 
and about what happens at the inner 
and outer disk radii. In some versions convection is `switched off' in
order to make a viscosity {\sl ansatz} work; in most, the
results depend on the number of grid points in the numerical code.
There are models in which some terms in the energy equation are dropped
even if they are of the same order as the terms left (the authors do
not hide this fact). Finally, some versions of the model, when applied to
particular classes of systems, require the help of tidal forces and
sometimes extremely low values of the parameter describing viscosity.
Of course some of these versions are the result of attempts to describe
particular systems.

In this respect the success of the DIM is not overwhelming. Attempts to
produce `standard' models of such `prototype' systems as the DN SS Cyg
and the SXT A0620--00 failed to produce results that would reproduce
fundamental properties of the prototype (such as the presence of
`inside--out' {\sl and} `outside in' outbursts in SS Cyg or the correct
rise--time and recurrence time in A0620--00).

So why does almost everybody (including the present authors) agree that the
DIM is the model describing DN and SXT outbursts? And why is it so
difficult to build a model that would incorporate the general
idea about the DIM?

The answer to the first question is that there is no competition.  At
the very beginning, when it was established that a dwarf nova is a
binary system in which a white dwarf accretes, through a disk matter
lost by a Roche--lobe filling secondary (see Warner 1995 and references
therein) star, there were some doubts as to where the outburst takes
place.  Soon it became clear that the site of the outburst is the
accretion disk (and not one of the stellar components) and two
possibilities were considered. The outburst could be due either to an
instability in the disk or to a (sudden) increase of the mass transfer rate
from the stellar companion of the compact object.  At first, the physical
reason for either  mechanism was not known, but after it was realized,
in the early eighties, that accretion disks in the parameter range
corresponding  to dwarf nova properties are thermally and viscously
unstable, the disk instability mechanism gained favour in the eyes of most
astrophysicists. The trigger for the supposed mass transfer instability
remained unknown and some observations, such as the behaviour of the
disk's outer radius during  outbursts, seemed to be inconsistent with
this mechanism. It was concluded, therefore, that the dwarf nova
phenomenon cannot be due to a sudden increase in mass transfer from the
secondary.

An attempt to revive the mass transfer model\cite{hkl86}
in the context of SXTs  was short lived because it was shown\cite{gh} that
this model, which used the X--ray irradiation of the secondary, cannot reproduce
observed time--scales.

\begin{figure} 
\centerline{\epsfig{file=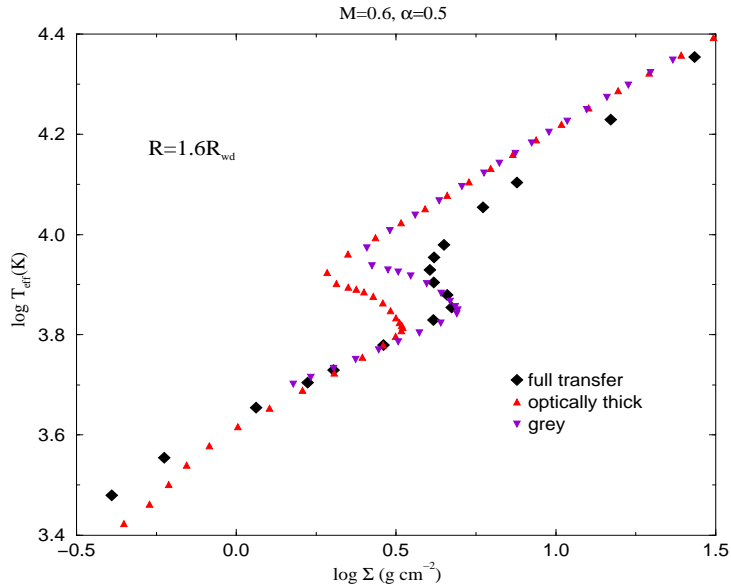,width=9.33cm,height=8cm}}
\vspace{10pt}
\caption{
$\Sigma$ -- $T_{\rm eff}$ curves for  $R=1.6 R_{\rm WD}$,
and $\alpha = 0.5$ for a $M=0.6 M_{\odot}$ white dwarf.
The vertical disk structure is calculated by using a full radiative
transfer code (Idan et al. 1998a) (diamonds), in the grey atmosphere 
(triangle down) and
optically thick (triangle up) approximation(HMDLH).
}\label{scurve}
\end{figure}

The (partial) answer to the second question is the subject of this review.

\section*{The viscosity problem}

The disk instability model of dwarf--nova outburst was formulated in
the early eighties by several groups of authors (for the history of the
early developments see Cannizzo\cite{cann93b},\cite{cann98a}).  
The various versions of the model used a
fundamental property of accretion discs:  in the range of temperatures
at which hydrogen recombines, the resulting dramatic reduction of
opacities makes the disk thermally and viscously unstable. This
property is usually shown on the $\Sigma$ -- $\nu \Sigma$ 
(surface density--(averaged) viscosity) plane:  at a
given radius, the disk thermal equilibria form  an $S$ - shaped
curve, the middle branch of which represents unstable solutions (see
Fig.~1 where  $S$ - curves are plotted on the surface density,
effective temperature ($T_{\rm eff}$) plane). If the rate at which
matter is fed into the disk corresponds, at some radius, to a locally
unstable state, the disk would have to oscillate between the lower,
cold, and the upper, hot, branch of the $S$-curve. The resulting limit cycle
is supposed to represent the dwarf nova outburst cycle.

From the very beginning it became clear that the existence of the
instability itself, as well as the disk behaviour, depend crucially
on properties of viscosity in the accretion flow. The viscosity is
described by the usual Shakura--Sunyaev prescription
$\nu = \alpha c_s H$
where $H$ is the half--thickness of the disk, $c_s$ the sound speed
and $\alpha$ the viscosity parameter, which in the original
formulation (Shakura \& Sunyaev 1973) was assumed
to be a constant $\leq 1$. It was first noted by
Smak\cite{smak84} that if $\alpha$ is kept constant, ``only very 
short--period, very low--amplitude variations could be produced".
To obtain light curves of dwarf novae Smak had to assume that $\alpha$
in a cold state is 4 times lower than $\alpha$ in the hot disk. This
jump in $\alpha$ is necessary because the jump in the temperature
itself, when moving from the cold to the hot branch is not big
enough to increase the viscosity to values that give the right
amplitudes and time-scales of dwarf nova outbursts.

As pointed out by Smak\cite{smak84} one should not expect the simple
$\alpha$ parameterization to represent all the complex phenomena due to
turbulent viscosity. Rather, once a reliable model
is found, observations of dwarf nova outbursts might be used to put
constraints on viscosity models. For example, it is thought that the
ratio  $\alpha_{\rm hot}/\alpha_{\rm cold}$ must be between 4 and 10 in
order to reproduce characteristic time-scales of DN outbursts. In most
cases values $\alpha_{\rm cold}\sim  0.01$ (see e.g. Livio and Spruit
1991) and $\alpha_{\rm hot}\sim0.1$ are believed to be required for models
to correspond to observations.

On the other hand, observations of quiescent DNs suggest
completely different values of $\alpha_{\rm cold}$. Quiescent DN disks
seem to be optically thin\cite{h93},\cite{whv}, whereas the DIM
predicts that between outbursts the disks are optically thick.  The
optical thickness is directly related to the value of $\alpha$ because
low $\alpha$'s imply high column densities. In a quiescent disk of the
DIM optical depths are high. Observed quiescent disks show the presence
of a Balmer jump in emission and brightness and colour temperature
distributions that are incompatible with an optically thick disk.
Attempts to fit observations with a simple model in which emission is
produced by a uniform, isothermal slab give $\alpha_{\rm cold} \gta
100$, which is unacceptably high. Modeling quiescent disk emission
using a radiative transfer code\cite{shwh91} gives $\alpha_{\rm
cold}\approx 0.5$\cite{ils}, which is in contradiction with the DIM.

It is not surprising that the main difficulties encountered in the DIM
are connected with viscosity. Despite the progress in the
understanding of the origin and nature of viscous turbulence
in accretion disks owing to the seminal work of Balbus and Hawley,
(see Balbus -- this volume)  we are still forced to use the $\alpha$
prescription. Only very recently have results
of numerical simulation been directly applied to disk model (e.g. Abramowicz,
Brandenburg \& Lasota 1996).

The first application of Balbus \& Hawley simulations to the problem of
dwarf nova outbursts questioned the validity of DIM itself\cite{gm98}.
Gammie and Menou\cite{gm98} calculated the magnetic Reynolds number
$R_{M}$ in a quiescent dwarf nova disk. The disk properties were
determined using the DIM code of Hameury et al. (1998) (which we will
discuss below), where, as usual, $\alpha_{\rm cold}\sim 0.01$. They
obtained $Re_{M} < 10^4$. According to the simulations with such low
values of $Re_{M}$ the instability creating the turbulence dies away.
It follows, if turbulence in accretion disks can result only from an
MHD instability\cite{bhs} (but see e.g. Yoshizawa \& Kato 1997), that there
would be no transport of angular momentum in quiescence, i.e. no
accretion.  This would mean that matter can accumulate only at
the outer disk rim. Since in quiescence there is no viscosity, the outburst
would have to be triggered by a MHD instability and not by a thermal
one. The DIM would have to be strongly modified.

It is too early however to try to modify drastically the `standard'
version of the DIM. First, observations clearly show that there is
accretion in quiescence (see also below). To this objection Gammie and Menou
answer that the observed emission could be due to a hot corona above the
cold `dead' disk. This a hypothesis that would have to be checked. Second,
one observes quite often `inside-out' outbursts which, it seems, it would
be difficult to get when matter is allowed to accumulate only in the
outer disk regions. 

It is interesting to note, however, that if $\alpha_{\rm cold}$ were $\sim 0.5$
then $R_{M} > 10^4$ and turbulence would be present the quiescent disk.
This value of $\alpha_{\rm cold}$ is the same as the one required by
observed emission from the disk\cite{ihlsh98}. Maybe this is not a
coincidence. 

In any case viscosity models are not yet ready to describe the vertical
stratification of viscosity or even its dependence on temperature and
other variables.
It is therefore not surprising that attempts were made to guess or to
postulate viscosity's functional dependence on variables such as 
temperature and radius. In the framework of the `$\alpha$--prescription'
one can only play with $\alpha$. The formula 
$\alpha = \alpha_0 \left(H/R \right)^n$
is rather popular nowadays.  It was first proposed by Meyer \&
Meyer--Hofmeister (1983) and used by Mineshige \& Wheeler (1989) in
their model of SXTs and by Cannizzo\cite{CCL},\cite{cann98b} also in
application to SXTs.  Ludwig, Meyer--Hofmeister \& Ritter (1994)
arrived at the conclusion that the $\alpha_{\rm hot}/\alpha_{\rm
cold}=4 - 10$ prescription  cannot produce `outside--in' outbursts and
suggest that $\alpha = \alpha_0 \left(H/R \right)^n$ might be necessary
to reproduce such types of outbursts. As shown, however, by Hameury et
al. (1998) (HMDLH; see also\cite{io92}) the lack of `outside--in'
outbursts in Ludwig et al. (1994) and similar calculations is mainly
due to the outer boundary condition that assumes a fixed outer disk
radius. When the radius is allowed to move during and after the
outburst (as it has in fact observed to do) there is no problem with
obtaining `outside--in' outbursts.

Another argument in favour of $\alpha = \alpha_0 
\left(H/R \right)^n$ is based on the shape of
light--curves during the decay from
outburst\cite{CCL},\cite{cann98a}. It is claimed that in order
to produce strictly exponential light--curves $\alpha$ has to be of the
form (2) with $n=1.5$. The value of $\alpha_0$ is supposed to depend on
the mass of the accreting body and to be $50$ for $M=10M_{\odot}$
(Mineshige \& Wheeler favor $\alpha_0\sim 10^3$ in their model of the
same systems).  There are several problems with this approach. First,
if $\alpha_0\sim 50$ then one encounters the same problem as when one
assumes $\alpha_0=const.$:  the jump in temperature on the $S$--curve
is not big enough to produce an outburst. Discovering this difficulty,
Cannizzo et al.\cite{CCL} decided to modify the physics leading to the
appearance of an $S$--curve.  Convection, as a cooling mechanism, plays
an important role in the accretion disk structure close to the local
surface density maximum on this curve. If one artificially, switches
off convection the resulting disk models produce the required amplitudes.
It is true that models of convection in accretion disks are far from 
being perfect, but discarding this physical mechanism just to obtain
the right shape of part of a light--curve seems  a bit far--fetched.

Especially since it is not certain how the form of $\alpha$ is related to
the shape of the {\sl observed} light--curves. First, it is not
clear that they are exactly exponential. According to Cannizzo (1998a)
the light curve of the SXT A0620--00 is rigorously exponential whereas
light curves of WZ Sge--type dwarf novae are `flat top' (power--like).
As shown, however, by Erik Kuulkers (private communication) the $V$ light
curves of A0620--00 and of the WZ Sge--type system AL Com  have {\sl
exactly} the same shape. In any case optical emission from SXT disks in
outburst is dominated by radiation due to X--ray reprocessing (Mc
Clintock \& van Paradijs 1995). The shape of the light curve during the
decay from outburst may be only very indirectly related to the $\alpha$
prescription (King \& Ritter 1997).

Finally, if $\alpha_0=50$ then the rise to outburst is much too slow
to agree with observations (Cannizzo 1998b). The rise to outburst
is due to the propagation of the heat front through the disk. This
front brings the disk from the quiescent cold state to the hot outburst
configuration. It propagates with the speed 
$v_{\rm fr} \approx \alpha c_s$
In the hot state $H/R \approx 0.01$\cite{cann98b} so that (with $\alpha_0=50$)
according to $\alpha\approx 0.05$ which is roughly ten times lower
than the value of $\alpha$ necessary to obtain the correct (i.e. observed)
rise-times to outburst. It is not surprising that Cannizzo (1998b) gets
rise-times 10 times too long.

As a conclusion one can state that arguments in favour of 
the $\alpha = \alpha_0  \left(H/R \right)^n$ formula are rather
weak.

\section*{Numerical codes}

Equations describing time--dependent accretion disks in the Keplerian,
geometrically thin ($H/R \ll 1$) approximation are equivalent to a
system of 4  non--linear, first order partial differential equations.
The rise to outburst is the result of the propagation through the disk
of a heating front; the decay from the maximum is, in general, when
illumination effects are neglected, the result of the propagation
through the disk of a cooling wave.  The problem is not mathematically
very complicated but if one wishes to resolve the fronts and describe
their propagation at scales differing by up to six orders of magnitude,
the required computer time may quickly become prohibitive. At first,
the number of (fixed) grid points in the DIM codes was less than 50,
and only recently (except for Mineshige\cite{m87})
Cannizzo\cite{CCL}\cite{cann98b} increased this number to
1000 (see HMDLH for discussion).

\begin{figure} 
\centerline{\epsfig{file=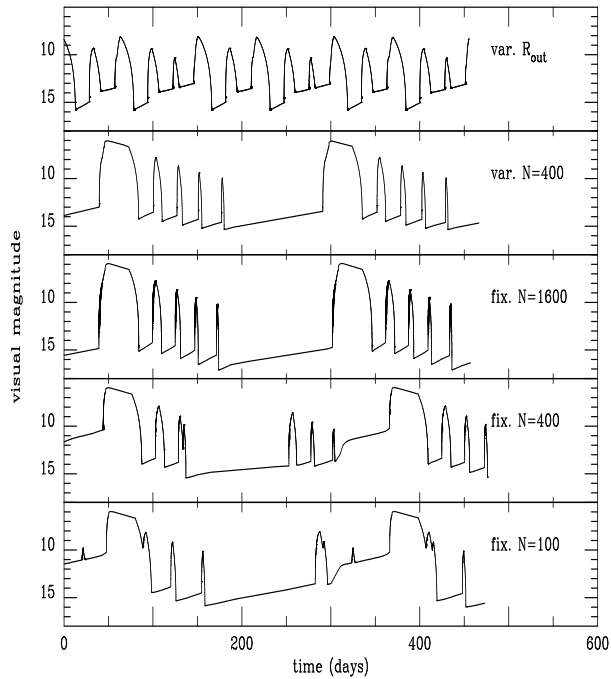, width=8cm,height=8.94cm}}
\vspace{10pt}
\caption{V magnitude light curves calculated using 
a fixed grid at various resolutions: (N=100, 400, 1600), adaptive grid 
with fixed (N=400) and variable $R_{\rm out}$ (N=1600).
}\label{lightcurves}
\end{figure}

Recently  a systematic study of the DIM
numerical problem was undertaken in several important articles.  
Cannizzo\cite{cann93b}, in a
pioneering article, showed that light curves (i.e. rise--times,
amplitudes, number of outbursts and recurrence times) strongly depend
on the number of grid points.  Unfortunately the number of points
chosen by Cannizzo himself (200 to 400) is, in general, not sufficient
to give physically reliable results (HMDLH). Moreover the character of
the outbursts (`inside-out') and the shape of the light-curves in his
models\cite{cann93b} are the result of keeping the outer disk radius
fixed. The same assumption was made in the important investigations of
Ludwig et al. (1994) and Ludwig and Meyer\cite{lm}. In this last
article it was shown that non--Keplerian effects arising from
steep gradients in the fronts can be neglected.

 All versions of the DIM codes suffered from various drawbacks such as
insufficient resolution of the grid, prohibitive computing time (for
explicit codes) and/or ill--adapted boundary conditions.  A new version
of the DIM code\cite{HMDLH} which is free of all these inconveniences has
been constructed by HMDLH. This implicit code uses an adaptive grid
and the size of the disk is allowed to vary. The number of grid points
can be as high as 4000 but this is not necessary.  HMDLH find that global
properties of transient discs can be addressed by codes using a high, but
reasonable, number of fixed grid points. However, the study of the
detailed physical properties of the transition fronts generally
requires resolutions which are out of reach of fixed grid codes. An
example in which fixed grid and adaptive grid calculations are compared
is shown on Figure 2. The HMDLH code can be considered as a continuation
of Smak's (1984) code and it has a lot in common with Ishikawa \&
Osaki (1992).

This new code will allow a systematic study of the DIM in the context
of both the dwarf--nova and X--ray transient outbursts. It has already been
used to model outbursts of WZ Sge and the rise to outburst of
GRO J1655-40 (see next Section).

\section*{Advection dominated accretion flows}

In the `standard' version of the DIM one assumes that the disk extends
all the way down to the central accreting object (or to the last stable
orbit in the case of a central black hole) and that  mass transfer
from the secondary is constant. Both these assumptions are difficult to
maintain when confronted with observations.

The DIM requires all  quiescent disks to be in the cold state (`on
the lower branch of the $S$--curve'). This implies that the accretion
rate throughout the disk must be lower than the critical value
corresponding to the local maximum surface density (the quiescent disk
is obviously {\sl not} in viscous equilibrium; if it were there would
be no outbursts). Since this critical accretion rate varies like
$R^{2.7}$ (see e.g. HMDLH) the resulting accretion rate onto the central
body could  be ridiculously small; in any case it is quite often
several orders of magnitude lower than the value deduced from observations
(see e.g. Lasota 1996). In the case of dwarf--novae whose quiescent
hard X--ray and UV emission is in contradiction with the DIM, Meyer and
Meyer--Hofmeister\cite{mm94} found  that the inner part of a quiescent
dwarf nova disk will evaporate and form a hot, tenuous accretion flow
which can emit the X--rays and UV required by the observations.

\begin{figure} 
\centerline{\epsfig{file=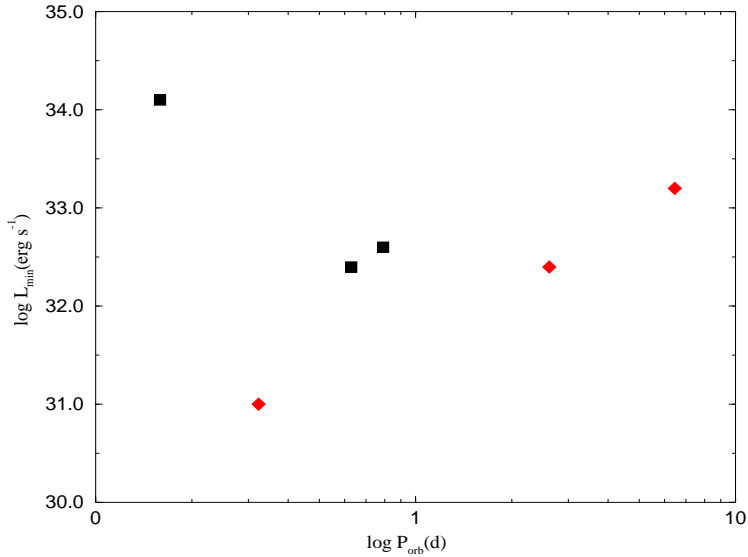,width=9.33cm,height=8cm}}
\vspace{10pt}
\caption{Quiescent luminosities of black-hole and neutron star soft X--ray
transients versus orbital period. Squares represent neutron star SXTs and
diamonds SXTs with confirmed ($M_{\rm BH}> 3 M_{\odot}$) black holes.
(Frank \& Lasota, 1997 unpublished)
}\label{quiel}
\end{figure}

In the case of quiescent black hole SXTs (BHSXTs) where the observed
quiescent X--ray flux cannot be explained in the framework of a
`standard' disk model, Narayan et al. \cite{nbm},\cite{nmy} showed that
an inner advection dominated accretion flow (ADAF) (see Menou, Quataert
\& Narayan 1998 for a recent review) provides a natural explanation of
the observational data.  Quiescent spectra of the three SXTs in which
X--rays in quiescence were detected are very well reproduced by an ADAF
model\cite{nbm},\cite{nmy},\cite{hlmn}.

The inner ADAF extends up to about $10^4$ Schwarzschild radii; the
outer accretion flow forms a dwarf--nova type disk subject to the
thermal--viscous instability which triggers the transient event. This
model was confirmed by the observation\cite{obrm} of the rise to
outburst of GRO J1655-40\cite{hlmn}.

In ADAFs most of the heat released by accretion is not radiated away in
the flow itself but advected to the central body. If this is a black
hole energy is lost forever below the event horizon; but if accretion is
onto a neutron star (or a white dwarf) the energy will eventually be
re-radiated from its surface. Therefore BHSXTs in quiescence should be
fainter than quiescent systems containing neutron stars. Of course this
statement is true if we compare systems with the {\sl same} mass
transfer rate.  Narayan, Garcia \& McClintock (1997)(see also
McClintock --these proceedings) compare outburst amplitudes versus
maximum luminosities of BHSXTs and neutron star systems. As predicted
by the ADAF model there is a clear difference between the two classes of
objects. The same clear difference is seen if one compares histograms
showing the distribution of the outburst amplitudes for the two classes
(Cannizzo\cite{cann98a} sees no difference because he includes in the
histogram {\sl lower limits} of BHSXT amplitudes).

Finally, one can (see Chen et al. these proceedings) try to compare
actual luminosities and not their ratios. The problem is to be able to
determine the mass transfer rate. If one assumes that it is related to
the orbital period one obtains the result shown on Fig.~3. It is clear
that the present data is not in contradiction with the ADAF model but
that much more observations are needed. One should also add that
quiescent neutron star SXTs could be dimmer than expected because of
the action of the propeller effect\cite{zh}.

\section*{Enhanced mass transfer}

As shown by Lasota, Narayan \& Yi (1996) in the ADAF + cold disk model
the outer disk can be (marginally) stable with respect to the thermal instability.
In this case the outburst would have to be triggered by an enhanced
mass transfer which would bring the outer disk into the unstable regime.

In any case it is observed that the mass transfer rate from the secondary varies on
various time--scales (see e.g. Warner 1995).  There is also evidence that
the mass transfer rate increases prior to superoutbursts of SU UMa dwarf--novae
and during some  `normal' outbursts (Smak 1996). This enhanced mass
transfer could be due to irradiation of the secondary.
It has been shown\cite{lhh},\cite{hlh} that if one wishes to avoid
using extremely low values ($\lta 10^{-4}$) of the viscosity parameter
required by the standard DIM one should consider a model in which the inner
part is evaporated {\sl and} mass--transfer is increased (by irradiation)
during the outburst.

Despite growing evidence to the contrary, it is still often asserted 
that dwarf nova outbursts can be explained by assuming that the mass 
transfer rate is constant
prior and during the outburst  (see e.g.
Cannizzo 1993a; Osaki 1996). The confrontation of such models with
observations does not, however, give satisfactory results. Clearly
the role of variations of the mass transfer rate in dwarf nova systems
requires more study.

\acknowledgments

We are grateful to Guillaume Dubus, Juhan Frank, Irit Idan and Kristen Menou 
for many enlightening discussions.

\endacknowledgments

\end{document}